\documentclass[twocolumn,prd,preprintnumbers,amsmath,amssymb,nofootinbib,superscriptaddress]{revtex4-1}
\usepackage[utf8]{inputenc}
\usepackage[dvipsnames]{xcolor}

\newcommand{\Lie}[1]{\ensuremath{{\cal L}_{#1}}}                          
\newcommand{\Poisson}[2]{\ensuremath{ \left\{ #1, #2\right\} }}
\newcommand{\scal}{\ensuremath{{\rm S}}}
\newcommand{\vect}{\ensuremath{{\rm V}}}
\newcommand{\tens}{\ensuremath{{\rm T}}}

\newcommand{\nablab}{\bar{\nabla}} 
\newcommand{\grad}{\ensuremath{\vec{\nabla}}}

\newcommand{\phib}{{\bar{\phi}}}

\newcommand{\AeST}{AeST~}
\newcommand{\lambdas}{\lambda_s}
\newcommand{\Mpc}{\ensuremath{{\rm Mpc}}}
\newcommand{\kpc}{\ensuremath{{\rm kpc}}}

\newcommand{\eps}{\epsilon}

\newcommand{\momentum}[3]{ {#1_{#2}}}
\newcommand{\Kcal}{{\cal K}}
\newcommand{\Lcal}{{\cal L}}

\newcommand{\Mpt}{\tilde{M}_p}

\newcommand{\Mcal}{{\cal M}}

\newcommand{\Fcal}{{\cal F}}

\newcommand{\Qcalb}{\bar{{\cal Q}}}
\newcommand{\Qcal}{{\cal Q}}
\newcommand{\Ycal}{{\cal Y}}

\newcommand{\Ccal}{{\cal C}}

\newcommand{\Hcal}{{\cal H}}
\newcommand{\Jcal}{{\cal J}}

\newcommand{\Gt}{\ensuremath{\tilde{G}}}
\newcommand{\GN}{\ensuremath{G_{{\rm N}}}}
\newcommand{\KB}{\ensuremath{K_{{\rm B}}}}

\begin{document}

\title{Aether scalar tensor theory: Linear stability on Minkowski space.}

\author{Constantinos Skordis}
\email{skordis@fzu.cz}
\affiliation{ CEICO, Institute of Physics of the Czech Academy of Sciences, Na Slovance 1999/2, 182 21, Prague, Czech Republic}             
\author{Tom Zlosnik}
\email{zlosnik@fzu.cz}
\affiliation{ CEICO, Institute of Physics of the Czech Academy of Sciences, Na Slovance 1999/2, 182 21, Prague, Czech Republic}                                                      
\date{\today}

\begin{abstract}  
We have recently proposed a simple relativistic theory which reduces to modified Newtonian dynamics for the weak-field quasistatic situations applied to galaxies, 
and to cosmological behavior as in the $\Lambda$CDM model, yielding a realistic cosmology in line with observations. 
A key requirement of any such model is that Minkowski space is stable against linear perturbations.
We expand the theory action to second
 order in perturbations on a Minkowski background and show that it leads to healthy dispersion relations involving propagating massive  modes
in the vector and the scalar sector. 
We use Hamiltonian methods to eliminate constraints present, demonstrate that the massive modes have Hamiltonian bounded from below
and show that a nonpropagating mode with a linear time dependence may have unbounded Hamiltonian for wave numbers $k< \mu$ and bounded otherwise. 
The scale $\mu$ is estimated to be $\lesssim \Mpc^{-1}$ so that the low momenta instability may only play a role on cosmological scales.
\end{abstract}

\maketitle

\section{Introduction}
The dark sector (DS) --dark matter and dark energy-- plays a pivotal role in cosmology and astrophysics. 
As yet, the evidence for the DS comes exclusively via its inferred contribution to the gravitational fields that known matter is observed to experience. Thus, it is possible that the phenomena of dark matter and/or dark energy may arise from a modification to the gravitational interaction.

Typically, theories of gravity different to general relativity (GR) introduce new degrees of freedom into the gravitational sector beyond the metric tensor present in GR \cite{Jain:2010ka,Clifton:2011jh}. While these degrees of freedom may have an important role to play in explaining aspects of the DS, it is crucial that they do not also introduce instabilities that are incompatible with observation. 

Observational constraints suggest that there exist regions of spacetime that can be approximated by highly symmetric solutions (for example geometry in the solar system can be described as a perturbed Minkowski spacetime, whereas the late universe on the largest scales can be described as perturbed de Sitter spacetime) and that these approximations persist for a proper time at least of the order $\tau_{s}$ (for example lower bounds on the age of the solar system or the period of $\Lambda$-domination in cosmology). 

It is vital then that new degrees of freedom do not introduce instabilities that grow on timescales $\tau_{i} \ll \tau_{s}$. To probe this question, one can consider 
the propagation of small perturbations to the aforementioned highly symmetric solutions. Classically, some theories of gravity allow perturbative modes that grow exponentially, 
where the timescale $\tau_{i}$ of growth may depend on basic parameters in the theories which can lead to significant constraints on their viability \cite{Seifert:2006kv,Seifert:2007fr}. 
Another possibility is that around some backgrounds, some perturbative modes can carry negative energy --either via wrong-sign kinetic terms (ghosts) 
or wrong-sign mass terms (tachyons). 
The former especially can signal pathological behavior in the quantum theory of these perturbations, signaling at the least that the background solution cannot be considered stable~\footnote{
We note that special cases have been constructed where the presence of a ghost does not lead to unstable behavior, see~\cite{Deffayet:2021nnt}.}.
If experimental constraints suggest that approximations to the background are long lived then this suggests that the theory of gravity 
in question is not healthy. Such considerations are therefore vital when considering the viability of a gravitational theory \cite{Boulware:1973my,Luty:2003vm,Cline:2003gs,Gorbunov:2005zk,DeFelice:2006pg,Rubakov:2008nh,Blas:2010hb,BeltranJimenez:2013btb,Chaichian:2014dfa,Langlois:2015cwa,BeltranJimenez:2019acz}.

We have recently proposed a relativistic theory which introduces additional fields in the gravitational sector
in order to account for the dark matter phenomenon~\cite{Skordis:2020eui}.
The theory depends on the metric tensor $g_{\mu\nu}$ but also introduces a unit timelike vector field $A_{\mu}$ --called Aether in the past~\cite{Jacobson:2000xp}--
 and a noncanonical shift-symmetric scalar field $\phi$ into the gravitational sector.
Hence, we refer to our proposal~\cite{Skordis:2020eui} as aether scalar tensor: \AeST.

 The new degrees of freedom in \AeST combine with the metric to produce modified Newtonian dynamics (MOND) phenomenology \cite{Milgrom1983a,BekensteinMilgrom1984} 
in the quasistatic, weak-field limit relevant to galaxies while accounting 
for precision cosmological data \cite{Planck:2018vyg} comparably well to the cold dark matter (CDM) 
paradigm \footnote{See \cite{Bekenstein1988,Sanders1997,Bekenstein2004,Sanders2005,NavarroVanAcoleyen2005,ZlosnikFerreiraStarkman2006,Sanders2007,Milgrom2009,BabichevDeffayetEsposito-Farese2011,DeffayetEsposito-FareseWoodard2011,Woodard2014,Khoury2014,Blanchet:2015sra,Hossenfelder2017,Burrage:2018zuj,Milgrom:2019rtd,DAmbrosio:2020nev} for alternative approaches to the construction of relativistic theories of gravity that contain MOND phenomenology.}.
The CDM-like cosmological behavior is unrelated to MOND but it is due to terms involving the new fields which have 
the same form as shift-symmetric $k$-essence and ghost condensate model~\citep{Scherrer2004,ArkaniHamedEtAl2003}.
This results in its cosmological energy density $\propto (1+z)^{3}$ plus small decaying corrections which makes
fitting large scale cosmological data possible.

Our goal is to study the linear stability of \AeST on a Minkowski background and to establish that
the theory is free of propagating ghost instabilities.
 In Sec \ref{Sec_theory} we introduce the theory in detail; in Sec \ref{Sec_Linear} we consider small fluctuations of the fields around a Minkowski background,
and expand the action to quadratic order. There, we also discuss gauge transformations and separately compute the dispersion relations
for tensor, vector, and scalar modes, determining at the same time the conditions on the theory parameters for these relations to be healthy.
 The case of scalar perturbations requires further treatment and in Sec \ref{Sec_Ham} we consider their Hamiltonian formulation.
We discuss our findings and their interpretation  in a cosmological setting in Sec \ref{Sec_discussion} and conclude in Sec \ref{Sec_conclusion}.

We use a metric signature $-+++$ and curvature conventions of Wald \cite{Wald:1984rg}.
We use brackets to denote antisymmetrization with the convention that $[A,B] = \frac{1}{2}(A B - B A)$.  

\section{The theory}
\label{Sec_theory}
\AeST depends on a metric $g_{\mu\nu}$ universally coupled to matter so that the Einstein equivalence principle is obeyed, a scalar field $\phi$ and a unit timelike vector field $A^\mu$. The action is 
\begin{widetext}
\begin{align}
S =&  \int d^4x \frac{\sqrt{-g}}{16\pi \Gt}\bigg\{ R  - 2 \Lambda
 - \frac{\KB}{2}  F^{\mu\nu} F_{\mu\nu} 
+ 2  (2-\KB) J^{\mu} \nabla_\mu \phi - (2-\KB) \Ycal
- \Fcal(\Ycal,\Qcal)
 - \lambda(A^\mu A_\mu+1)\bigg\}  + S_m[g]
\label{NT_A_action}
\end{align}
\end{widetext}
where $g$ is the metric determinant, $\nabla_\mu$ the covariant derivative compatible with $g_{\mu\nu}$,   $R$  is the Ricci scalar, $\Lambda$ is the cosmological constant,
$\Gt$ is the bare gravitational strength, $\KB$ is a constant and $\lambda$ is a Lagrange multiplier imposing the unit timelike constraint on $A_{\mu}$. 
The matter action $S_m$ is  assumed not to depend explicitly on $\phi$ or $A^\mu$.
\AeST has a function $\Fcal(\Ycal,\Qcal)$ which depends on the scalars $\Qcal = A^\mu \nabla_\mu \phi$ and $\Ycal= (g^{\mu\nu} + A^{\mu} A^{\nu})\nabla_\mu \phi \nabla_\nu \phi$, while
$J^\mu = A^\nu \nabla_\nu A^\mu$ and $F_{\mu\nu} = 2\nabla_{[\mu} A_{\nu]}$. 
The function $\Fcal$ is subject to conditions so that the  cosmology of \AeST is compatible with $\Lambda$CDM on
FRLW spacetimes and a MOND limit emerges in quasistatic situations.

On a flat FLRW background the metric takes the form $ds^2 = -dt^2 + a^2 \gamma_{ij} dx^i dx^j$ where $a(t)$ is the scale factor and $\gamma_{ij}$ is a flat spatial metric.
The vector field reduces to $A^\mu = ( 1, 0, 0, 0)$ while $\phi \rightarrow \phib(t)$ leading to $\Qcal\rightarrow \Qcalb = \dot{\phib}$ and  $\Ycal \rightarrow 0$,
so that we may define $\Kcal(\Qcalb) \equiv -\frac{1}{2} \Fcal(0,\Qcalb)$.
We require that $\Kcal(\Qcalb)$ has a minimum at $\Qcal_0$ (a constant) so that we may expand it as $\Kcal = \Kcal_2\left(\Qcalb - \Qcal_0\right)^2 + \ldots$, where the $(\ldots)$ denote 
higher terms.
This condition leads to $\phib$ contributing energy density scaling as dust $\sim a^{-3}$ akin to~\cite{Scherrer2004,ArkaniHamedEtAl2003},  plus small corrections which tend 
to zero when $a\rightarrow \infty$. In principle, $\Kcal$ could be offset from zero at the minimum $\Qcal_0$, i.e. $\Kcal(\Qcal_0) = \Kcal_0$,
 however, such an offset can always be absorbed into the cosmological constant $\Lambda$ and thus we choose $\Kcal_0 = 0$ by convention, implying the same on the parent function $\Fcal$.

In the quasistatic weak-field limit we may set the scalar time derivative to be at the minimum $\Qcal_0$, as is expected to be the case in the late universe.
This means that we may expand $\phi = \Qcal_0 t + \varphi$. Moreover, in this limit $\Fcal \rightarrow  (2-\KB) \Jcal(\Ycal)$, with $\Jcal$ defined appropriately as
$\Jcal(\Ycal) \equiv \frac{1}{2 -\KB} \Fcal(\Ycal,\Qcal_0)$.
It turns out that MOND behavior emerges if $\Jcal \rightarrow \frac{2\lambdas}{3(1+\lambdas) a_0} |\Ycal|^{3/2}$ where
$a_0$ is Milgrom's constant and $\lambdas$ is a constant which is related to the Newtonian/GR limit. Specifically, there are two ways that GR can be restored: (i) screening 
and (ii) tracking.
In the former, the scalar is screened at large gradients $\grad \varphi$, where $\grad \leftrightarrow \grad_i$ is the spatial gradient on a flat background $\gamma_{ij}$,
and in the latter, $\lambdas \varphi$ becomes proportional to the Newtonian potential, leading to an effective Newtonian constant 
\begin{equation}
\GN = \frac{1 + \frac{1}{\lambdas}}{ 1 - \frac{\KB}{2} } \Gt.
\end{equation}
Screening may be achieved either through terms in $\Jcal \sim \Ycal^p$ with $p>3/2$ or through Galileon-type terms which must be added to \eqref{NT_A_action}. Either way, for our purposes in this article,
we may model screening as $\lambdas \rightarrow \infty$.

\section{Linear perturbations around Minkowski space}
\label{Sec_Linear}
\subsection{Perturbative setup}
We are interested in spacetime regions which are well approximated by weak gravitational fields modeled as fluctuations on a Minkowski background $\eta_{\mu\nu}$ and that these regions
exist in the late universe where the time derivative of the background field has settled in its minimum $\Qcal_0$, i.e. $\dot{\phib} \rightarrow \Qcal_0$.
In addition, the size of these regions is taken to be much smaller than the size of the current cosmological horizon so that we may safely ignore the cosmological constant.

We expand the metric as $g^{\mu\nu} = \eta^{\mu\nu} - h^{\mu\nu}$, where $\eta_{00} = -1$ and $\eta_{ij} = \gamma_{ij}$, the vector field~\footnote{
 Strictly speaking, to satisfy the Lagrange constraint we need $A_0$ to second order, i.e. $A_0 =  -1 + \frac{1}{2} h^{00}   - \frac{3}{8} (h^{00})^2 - \frac{1}{2} |\vec{A}|^2  -  h^{0i} \vec{A}_i$
and similarly for $A^0$. However, for all the other terms in \eqref{NT_A_action}, it is sufficient to expand $A_0$  and $A^0$ to first order. }
  as $A_\mu = ( -1 + \frac{1}{2} h^{00} , \vec{A}_i)$ and the scalar as $\phi = \Qcal_0 t + \varphi$. 
 Thus our degrees of freedom are the metric perturbation $h^{\mu\nu}$, vector field perturbation $\vec{A}_i$ (only its 3-dimensional part remains free) 
and the scalar field perturbation $\varphi$, all of which are in general functions of both space and time.
We raise/lower spatial indices with the spatial metric $\gamma_{ij}$, i.e. $\vec{A}^i = \gamma^{ij} \vec{A}_j$ and 
set $|\vec{A}|^2 = \vec{A} \cdot \vec{A} = \vec{A}_i \vec{A}^i$ (and use similar notation for other  spatial vectors).

\subsection{Gauge transformations}
Our perturbative variables are amenable to gauge tranformations generated by a vector field $\xi^\mu$. Generally, for a tensor ${\bf A}$, its perturbation transforms as
$\delta {\bf A} \rightarrow \delta {\bf A} + \Lie{\xi} \bar{{\bf A}} $. Usually, on Minkowski space only the metric has a nonzero background value ($\eta_{\mu\nu}$), so that other fields besides
the metric perturbation are gauge invariant on such a background; this is  typical of dark fields, i.e. additional degrees of freedom which contribute to the energy density but do not
mix with the metric perturbation through gauge transformations of this kind. 
In our case, however, both the vector field and the scalar field have nonzero background value: $\bar{A}_\mu = (-1,0,0,0)$ and $\phib = \Qcal_0 t$, hence, their perturbations do
transform. Specifically, parametrizing $\xi^\mu$ as $\xi^\mu = (\xi_T, \vec{\xi}^i)$, we have the usual metric gauge transformations
\begin{align}
h_{\mu\nu} \rightarrow h_{\mu\nu} + \nablab_\mu \xi_\nu + \nablab_\nu \xi_\mu 
\end{align}
where $\nablab_\mu$ is the covariant derivative associated with the Minkowski metric $\eta_{\mu\nu}$.
In $3+1$ form the above transformations are explicitly given as
\begin{align}
h_{00} \rightarrow h_{00} - 2 \dot{\xi}_T  
\\
h_{0i} \rightarrow h_{0i} + \dot{\vec{\xi}}_i - \grad_i \xi_T
\\
h_{ij} \rightarrow h_{ij} + \grad_i \vec{\xi}_j + \grad_j \vec{\xi}_i.
\end{align}
The perturbations $\vec{A}$ and $\varphi$ transform as
\begin{align}
\vec{A} \rightarrow \vec{A}  -  \grad \xi_T
\\
\varphi \rightarrow \varphi +  \Qcal_0 \xi_T.
\end{align}
Notice how the vector field transformation has the same form as gauge transformations in electromagnetism, however, the generator here is also a diffeomorphism.

With these gauge transformations at hand we can create the following gauge-invariant variables:
\begin{align}
 \{ \grad \varphi +  \Qcal_0  \vec{A}, \dot{\vec{A}} - \frac{1}{2} \grad h_{00}  , \dot{\varphi} + \frac{1}{2} \Qcal_0 h_{00}  \}
\end{align}
Hence, the fields $\varphi$ and $\vec{A}_i$ nontrivially mix with the metric perturbation through $\xi_T$.

\subsection{The second order action}
 Our aim is to then expand the action \eqref{NT_A_action} to second order in these fields.
With these considerations, and having in mind the discussion in the previous section, we then expand the function $\Fcal$ as
\begin{equation}
\Fcal =   (2-\KB)\lambdas \Ycal  - 2 \Kcal_2  \left(\Qcal - \Qcal_0\right)^2 + \ldots
\label{Fcal_exp}
\end{equation}
since  $\bar{\Fcal}(0,\Qcal_0) = 0$ by convention and $\frac{\partial \bar{\Fcal}}{\partial \Qcal}\big|_{ \{0,\Qcal_0 \} } = 0$ at the minimum. The 
terms denoted by $(\ldots)$ are higher order terms which do not contribute to the second order action. We particularly note that one of these is the MOND-type term $\sim |\Ycal|^{3/2}$ 
as discussed in the previous section. This term does not contribute to the second order action but we return to it in the discussion section.

As an example, consider  the function 
\begin{align}
\Fcal =&
- 2 \Kcal_2 (\Qcal - \Qcal_0)^2
+ \lambdas \bigg\{\Ycal - 2 a_0 (1+\lambdas) \sqrt{\Ycal} 
\nonumber
\\
&
+2 (1+\lambdas)^2a_0^2 \ln\left[1  +  \frac{\sqrt{\Ycal}}{ (1+\lambdas)a_0}\right]\bigg\} 
\end{align}
In the large $\Ycal$ limit, the expansion \eqref{Fcal_exp} is recovered and the leading correction is $\sim \sqrt{\Ycal}$, while in the small $\Ycal$ limit,
 the expansion is consistent with \eqref{Fcal_exp} upon setting $\lambdas=0$ and the leading correction is the MOND term $\frac{2\lambdas}{3(1+\lambdas) a_0} |\Ycal|^{3/2}$
Notice the presence of $\lambdas$ as a relic of its influence on the observed value of Newton's constant in strong gravity regimes.

Expanding \eqref{NT_A_action} to second order
leads to
\begin{widetext}
\begin{align}
S =&  \int d^4x \bigg\{ - \frac{1}{2} \nablab_\mu h \nablab_\nu h^{\mu\nu}
+ \frac{1}{4}  \nablab_\rho h \nablab^\rho h
+ \frac{1}{2} \nablab_{\mu} h^{\mu\rho}  \nablab_\nu h^\nu_{\;\;\rho}
- \frac{1}{4} \nablab^\rho h^{\mu\nu}  \nablab_\rho  h_{\mu\nu}
+ \KB |\dot{\vec{A}} - \frac{1}{2} \grad h^{00}|^2
- 2\KB \grad_{[i} A_{j]} \grad^{[i} A^{j]}
\nonumber
\\
&
+  \left(2 - \KB\right)\left[  2(  \dot{\vec{A}}  -  \frac{1}{2}  \grad h^{00} ) \cdot ( \grad \varphi + \Qcal_0 \vec{A} ) 
- ( 1 + \lambdas )  |\grad\varphi + \Qcal_0 \vec{A}|^2
\right]
+2 \Kcal_2  \left|\dot{\varphi} +  \frac{1}{2}\Qcal_0 h^{00} \right|^2
 +  \frac{1}{\Mpt^2} T_{\mu\nu} h^{\mu\nu}
 \bigg\}
\label{Minkowski_action}
\end{align}
\end{widetext}
where for convenience we have rescaled the action $S \rightarrow 16\pi \tilde{G}S$. We have also omitted the determinant $\sqrt{\gamma}$ in the measure since
we are dealing with integrals on Minkowski spacetime, but can be understood to be present in all integrations.

We decompose the fields into scalar, vector and tensor harmonics as
\begin{align}
h_{00} =&  -2\Psi 
\\
h_{0i} =&  - \grad_i \zeta -  W_i
\\
h_{ij} =&  -2\Phi \gamma_{ij} +   D_{ij} \nu + 2\grad_{(i}V_{j)} + H_{ij} 
\\
\vec{A} =& \grad \alpha + \vec{\beta}
\end{align}
where $D_{ij} = \grad_i\grad_j-\frac{1}{3}\gamma_{ij}\grad^2$ is a traceless derivative operator. The modes $\vec{W}$, $\vec{V}$ and $\vec{\beta}$ are pure vector modes, that is, they are 
transverse: $\grad \cdot \vec{W} = \grad \cdot \vec{V} = \grad \cdot \vec{\beta} = 0$, while the mode $H_{ij}$ is a pure tensor mode, that is, transverse and traceless: $\grad_i H^i_{\;\;j} = H^i_{\;\;i} = 0$.

The matter stress-energy tensor $T_{\mu\nu}$ is likewise decomposed as
\begin{align}
T_{00} =&  \rho
\\
T_{0i} =&  \grad_i \theta +  p_i
\\
T_{ij} =&  P \gamma_{ij} +   D_{ij} \Sigma^{(\scal)} + 2 \grad_{(i} \Sigma^{(\vect)}_{j)} + \Sigma^{(\tens)}_{ij}
\end{align}
where the scalar modes are the matter density $\rho$, momentum divergence $\theta$, pressure $P$ and scalar shear $\Sigma^{(\scal)}$, 
the vector modes are the matter vorticinal momentum density $p_i$ and vector shear $\Sigma^{(\vect)}_i$, such that $\grad \cdot \vec{p} = \grad\cdot \vec{\Sigma}^{(\vect)} = 0$, 
and the tensor mode is the tensor shear $\Sigma^{(\tens)}_{ij}$, such that $\grad_i \Sigma^{(\tens)i}_{\;\;\;\;j} = \Sigma^{(\tens)i}_{\;\;\;\;i} = 0$.

With this decomposition, the second order action splits into three distinct parts: one for the scalar modes $S^{(\scal)}$, one for the vector modes $S^{(\vect)}$ and one for the tensor modes $S^{(\tens)}$. 
We consider each of these three one by one.

\subsection{Tensor modes}

The perturbations to fields $A_\mu$ and $\phi$ do not contribute any tensor mode components and so  the tensor mode action takes the form:
\begin{align}
S^{(\tens)} &=  \int d^{4}x  \bigg\{\dot{H}^{ij}\dot{H}_{ij}- \grad^{k}H^{ij}\grad_{k}H_{ij}
 + 32 \pi \Gt  \Sigma^{(\tens)}_{ij} H^{ij} \bigg\}
\end{align}
This corresponds to the action for tensor modes present in general relativity, a result consistent with the earlier, 
more general calculation that tensor modes in the superclass of theories of which \eqref{NT_A_action} is a special subset, propagate at the speed of light~\cite{SkordisZlosnik2019}.
 
\subsection{Vector modes}

We now consider vector modes, which are described  by the action
\begin{align}
S^{(\vect)} =&  \int d^4x  \bigg\{
-  \frac{1}{2} \left(\dot{V}_{i} + W_i\right)\grad^2\left(\dot{V}^{i}  + W^i\right)
 \nonumber 
\\
& +  K_{B}\left[ |\dot{\vec{\beta}}|^2 - \grad_i \beta_j \grad^i \beta^j -  \Mcal^2  |\vec{\beta}|^2 \right]
 \nonumber 
\\
&
+ 16\pi \Gt  \left(    \vec{p} \cdot \vec{W} - \Sigma^{(\vect)}_i \grad^2 V^i 
\right)
 \bigg\} 
\end{align}
where
\begin{equation}
 \Mcal^2 = \frac{(2-\KB)(1 + \lambdas )  \Qcal_0^2}{\KB}
\label{Mass_of_modes}
\end{equation}
The field $\vec{\beta}$  decouples from the metric fields $\vec{V}$ and $\vec{W}$ and describes two massive degrees of freedom with mass $\Mcal$. 
Clearly then we must require $\KB>0$ to avoid ghosts and gradient instabilities. The mass term $\Mcal$
 is also nontachyonic if both $0<\KB<2$ and $\lambdas>-1$. 
Hence,  stability considerations for the vector modes imply the following constraints on the parameter space of~\AeST:
\begin{align}
 0<\KB<2, \quad \lambdas > -1.
  \label{vector_constraints}
\end{align}
Notice that to this order, the vector modes $\vec{\beta}$ do not couple to matter and thus they are not expected to be generated by sources to leading order.

\subsection{Scalar modes}

We now consider scalar perturbations. Considering only scalar modes in \eqref{Minkowski_action} and after some integrations by parts we find the action $S^{(\scal)}$ :    
\begin{widetext}
\begin{align}
S^{(\scal)} =& \int d^4x  \bigg\{
6 \left( \frac{1}{6} \grad^2\dot{\nu}     - \dot{\Phi} \right) \left( \frac{1}{6} \grad^2\dot{\nu}     + \dot{\Phi} \right)
+ 4 \left(  \frac{1}{6} \grad^2\dot{\nu}  +  \dot{\Phi} \right) \grad^2\zeta
+ 2 |\grad\Phi|^2 - \frac{2}{3}\Phi \grad^4\nu 
+ 4\left( \grad^2\Phi  + \frac{1}{6} \grad^4\nu\right) \Psi
\nonumber
\\
&
+ \frac{1}{18}  |\grad (\grad^2\nu)|^2
+2 \Kcal_2 \dot{\varphi}^2 
 -  4 \Kcal_2   \Qcal_0  \dot{\varphi}  \Psi
 +  2 \Kcal_2  \Qcal_0^2  \Psi^2 
 + \KB  |\grad (\dot{\alpha} + \Psi)|^2 
+ 2 \left(2 - \KB\right)   \grad\left( \dot{\alpha}  +   \Psi  \right) \cdot \grad \chi
\nonumber 
\\
&
- \left(2-\KB\right)\left(1 + \lambdas \right) |\grad\chi|^2  
- 16\pi \Gt \rho \Psi
- 16\pi \Gt  \grad^2 \theta \, \zeta
- 48\pi \Gt  P  \Phi
+ \frac{16\pi \Gt }{3} \grad^4\Sigma   \, \nu
\bigg\}
\label{realspacescalars}
\end{align}
\end{widetext}
where we have defined the gauge-invariant variable $\chi$ as
\begin{equation}
\chi \equiv \varphi + \Qcal_0 \alpha
\end{equation}
that will be shown to play a prominent role in what follows.

Setting scalar matter sources to vanish and moving to Fourier space we have
\begin{widetext}
\begin{align}
S^{(\scal)} =& \int dt \frac{d^3k}{(2\pi)^3}  \bigg\{
    - 6 |\dot{\Phi}|^2
+   \frac{1}{6} k^4 |\dot{\nu}|^2
+ 2 k^2  \left[ 
\left(  \frac{1}{6} k^2\dot{\nu}  -  \dot{\Phi} \right) \zeta^*
+ c.c. 
\right]
+ 2 k^2 |\Phi - \frac{1}{6} k^2 \nu |^2
+2 \Kcal_2 |\dot{\varphi}  -  \Qcal_0 \Psi|^2 
 + \KB k^2  |\dot{\alpha} + \Psi|^2 
\nonumber
\\
&
- 2k^2 \left[ \left( \Phi  - \frac{1}{6} k^2\nu\right) \Psi^*  + c.c.
\right]
+ \left(2 - \KB\right)   k^2 \left[ \left( \dot{\alpha}  +   \Psi  \right) \chi^* 
+ c.c.
\right]
- \left(2-\KB\right)\left(1 + \lambdas \right) k^2 |\chi|^2  
\bigg\}
\label{scalar_fourier_action}
\end{align}
\end{widetext}
where fields in \eqref{scalar_fourier_action} have a subscript $\vec{k}$ to explicitly 
show their $k$ dependence as they are 
the Fourier modes of those in (\ref{realspacescalars}) and (c.c) means complex conjugate.

We now find the normal modes. It is sufficient to work in the Newtonian gauge by setting $\nu = \zeta = 0$.
We set the time dependence of all perturbations to $e^{i\omega t}$ 
 and rewrite \eqref{scalar_fourier_action} as $ \int dt \int \frac{d^{3}k }{(2\pi)^3} Z^\dagger U  Z + (h.c) $, where 
$Z = \{ \Psi, \Phi, \alpha,\varphi\}$ and $U$ is a $4\times 4$ matrix of coefficients which depend on $\omega$, $k$ and the other \AeST parameters.
The determinant of $U$ is found to be
\begin{align}
\det U =&  4k^6 \omega^2 \bigg\{  (2-\KB) \left[ (2 + \KB \lambdas) k^2 + 2 \Kcal_2  \Qcal_0^2 (1+\lambdas) \right] 
\nonumber 
\\
& - 2 \Kcal_2 \KB \omega^2 \bigg\}
\end{align}
so setting $\det U=0$ gives the two  dispersion relations 
\begin{align}
 \omega^2 =& 0
\\
 \omega^2 =& c_s^2 k^2  + \Mcal^2
\label{eq_dispersion}
\end{align}
where the scalar speed of sound is
\begin{equation}
c_s^2 = \frac{(2-\KB)}{\Kcal_2 \KB}  (1 + \frac{1}{2} \KB \lambdas)  
\label{speed_of_sound}
\end{equation}
We notice that the first mode does not lead to a propagating wave but rather to a mode evolving as $\sim A_0 + B_0 t$ where $A_0$ and $B_0$ are $k$-dependent constants.
Interestingly also, the second mode is massive with the same mass as the vector mode $\vec{\beta}$.

Positivity of $c_s^2$ implies further stability conditions in addition to the ones found above for the vector modes. Specifically, 
since from \eqref{vector_constraints} we have $\lambdas>-1$, then $1 + \frac{1}{2} \KB \lambdas>0$ leading to the condition
\begin{align}
\Kcal_2 > 0.
\label{scalar_constraints}
\end{align}

The other two would-be normal modes are nondynamical, i.e. they have no kinetic term and do not contribute a term involving $\omega$.  This signifies the presence of constraints
which are revealed through Hamiltonian analysis. We proceed to do so now as it also sheds more light on the $\omega=0$ mode.

\section{Hamiltonian formulation of scalar modes}
\label{Sec_Ham}
We now move to the Hamiltonian description of scalar modes which serves a double purpose. 
It allows us to de-constrain the system by removing the redundant gauge and nondynamical degrees of freedom and 
further investigate the significance of the $\omega=0$ normal mode. Starting from \eqref{scalar_fourier_action}, notice that out of the six fields, two ($\Psi$ and $\zeta$) do not contain time derivatives.
We determine the canonical momenta (see  Appendix \ref{Def_Fourier_Ham})
for the other four which are found to be
\begin{align}
\momentum{P}{\Phi}{k} =& -4 \left( 3 \dot{\Phi} +  k^2   \zeta \right),
\label{Momentum_P_Phi}
\\
\momentum{P}{\nu}{k} 
 =& \frac{1}{3}  k^4 \left(\dot{\nu} + 2 \zeta\right),
\\
\momentum{P}{\chi}{k}  =& 4  \Kcal_2 \left[\dot{\chi}  - \Qcal_0(\dot{\alpha} +   \Psi) \right],
\\
\momentum{P}{\alpha}{k}  =& 
 -4 \Kcal_2 \Qcal_0 \dot{\chi}  
 + 2 \left( \KB k^2  + 2 \Kcal_2 \Qcal_0^2 \right)  \left(\dot{\alpha} + \Psi\right)
\nonumber
\\
& 
+ 2\left(2 - \KB\right)   k^2 \chi,
\end{align}
and where we opted to use $\chi$ rather than $\varphi$ as the dynamical variable.
Performing a Legendre transformation we then find the Hamiltonian density as
\begin{widetext}
\begin{align}
\Hcal =&  
-\frac{1}{24} |P_\Phi|^2
+ \frac{3}{2k^4} |P_\nu|^2
+ \frac{1}{8\Kcal_2} |P_\chi|^2
+ \frac{1}{4k^2 \KB}   | P_\alpha + \Qcal_0 P_\chi |^2 
- 2 k^2   |\Phi - \frac{1}{6}k^2 \nu|^2
+ \frac{2-\KB}{\KB}  k^2 (2+\KB \lambdas) |\chi|^2
\nonumber
\\
&
- \frac{2-\KB}{2\KB} \left[(P_\alpha + \Qcal_0 P_\chi ) \chi^*  + ( P_\alpha^* + \Qcal_0 P_\chi^*   )    \chi\right]
+ C_\Psi  \Psi^*
+ C_\Psi^*  \Psi
+ C_\zeta^*  \zeta
+ C_\zeta  \zeta^* 
\label{scalar_Hamiltonian}
\end{align}
\end{widetext}

Since the  variables $\Psi$ and $\zeta$ are not dynamical, their function is to act as Lagrange multipliers imposing the constraints
\begin{align}
C_\Psi \equiv & 2 k^2   \Phi - \frac{k^4}{3}    \nu - \frac{1}{2} P_\alpha 
\approx 0
\\
C_\zeta \equiv &  - P_\nu - \frac{k^2}{6}  P_\Phi   
\approx 0
\end{align}
which essentially cast $\nu$ and $P_\nu$ as functions of the other variables. As usual we use the symbol $\approx$ to denote weakly vanishing constraints (those that vanish only on-shell)~\cite{Dirac:1958sq}.
Notice also that the variable $\alpha$ is cyclic, therefore its canonical momentum $P_\alpha$ is conserved and is an integral of motion.


%
 We require that the constraints are preserved by time evolution according to the Hamiltonian $H = \int \frac{d^{3}k}{(2\pi)^3}\Hcal $. We define the Poisson brackets on phase space as 
\begin{align}
\Poisson{f}{g}
 &=  (2\pi)^{3}\int d^{3}k \left[ \sum_I  \left(\frac{\delta f}{\delta X^I} \frac{\delta g}{\delta P^*_{X^I}} -  \frac{\delta g}{\delta X^I} \frac{\delta f}{\delta P^*_{X^I}} \right)
  \right]
\end{align}
where $I$ runs over $\{\Phi, \nu, \chi, \alpha\}$.  The time evolution of a variable $f$ is
\begin{align}
\dot{f} &= \Poisson{f}{H},
\end{align}
so we have
\begin{align}
\dot{C}_{\Psi} & = C_{\zeta},
\\
\dot{C}_{\zeta} &=0.
\end{align}
Hence, the constraints are preserved by time evolution on-shell.
 Therefore as one might expect, the stability of the primary constraints in the absence of gauge fixing does not create new constraints. 
Having ensured the stability of constraints in the Hamiltonian, we can now simplify the system by employing gauge fixing.

 In the Hamiltonian formulation, primary first-class constraints generate gauge transformations. The infinitesimal change of a phase space quantity $f$ under this gauge transformation 
generated by the constraint $C_{I}$ is given by:
\begin{align}
\Delta f = \Poisson{f}{C_{I}^*[\eps_{I}]}.
\end{align}
where we have introduced the smearing $C^{*}_{I}[\eps_{I}]$ of a constraint $C^{*}_{I}$ with test function $\eps_{I}$ defined as 

\begin{align}
	C^{*}_{I}[\eps_{I}] \equiv \int \frac{d^3k}{(2\pi)^{3}} \eps_{I,\vec{k}}C^{*}_{I,\vec{k}}
\end{align}
Consider the following gauge transformations generated by the constraints $C_{\zeta}$ and $C_{\Psi}$:
\begin{align}
\Delta \nu &= \Poisson{\nu}{ C_{\zeta}^*[\eps_{\zeta}]} = - \eps_{\zeta} 
\\
\Delta P_{\nu} &= \Poisson{P_{\nu}}{ C_{\Psi}^*[\eps_{\Psi}]} =  \frac{1}{3}k^{4}\eps_{\Psi}
\end{align}
Thus, we may set $\nu$ and $P_{\nu}$ to zero by a gauge transformation by choosing $\eps_{\zeta} = \nu$ and $\eps_{\Psi}= -\frac{3}{k^{4}}P_{\nu}$.
We then check what constraints are placed on the Lagrange multipliers $\zeta,\Psi$ by this gauge fixing. We invoke two new gauge fixing constraints:
\begin{align}
G_{\nu}  &\equiv \nu \approx 0 \\
G_{P_{\nu}} &\equiv  P_{\nu} \approx 0
\end{align}
and find
\begin{align}
\Poisson{G_{\nu}}{H} &= \frac{3}{k^{4}}G_{P_{\nu}} -2\zeta
\\
\Poisson{G_{P_{\nu}}}{H} &= \frac{2}{3}k^{4}\left(\Psi-\Phi\right) + \frac{1}{9}k^{6}G_{\nu}
\end{align}
Therefore the following gauge restrictions are placed on the Lagrange multipliers: $\zeta = 0$ and $\Psi = \Phi$. 
We recognize these conditions, respectively, as a restriction to the conformal Newtonian gauge and the content of the Einstein equation here dictating equality between metric potentials in this gauge. 
We may adopt these conditions alongside the constraints $G_{\nu}$, $G_{P_{\nu}}$ in the Hamiltonian \eqref{scalar_Hamiltonian} and the primary constraints, yielding in addition 
\begin{align}
P_\Phi &\approx 0
\\
\Phi &\approx \frac{1}{4k^2} P_\alpha
\end{align}
so that the deconstrained Hamiltonian density is
\begin{align}
\Hcal^{({\rm Dec})} =&  
 \frac{1}{8\Kcal_2} |P_\chi|^2
+ \frac{1}{4k^2 \KB}   | P_\alpha + \Qcal_0 P_\chi |^2 
- \frac{1}{8 k^2}   |P_\alpha|^2
\nonumber
\\
&
- \frac{2-\KB}{2\KB} \left[(P_\alpha + \Qcal_0 P_\chi ) \chi^*  + ( P_\alpha^* + \Qcal_0 P_\chi^*   )    \chi\right]
\nonumber
\\
&
+ \frac{2-\KB}{\KB}  k^2 (2+\KB \lambdas) |\chi|^2.
\label{scalar_Hamiltonian_deconstrained}
\end{align}
The Hamiltonian density $\Hcal^{({\rm Dec})}$ is free of constraints but its form remains rather complicated. 
We can make an additional simplification by making a canonical transformation to canonical pairs $(P_{X},X)$, $(P_{Y},Y)$ defined via
\begin{widetext}
\begin{align}
\chi =&  \sqrt{\frac{ \KB k^2  + (2-\KB) \mu^2 }{\KB(2-\KB) }}\frac{\Qcal_0}{\mu k} \, X  +  \frac{1}{2} \frac{P_Y}{ (2 +\KB \lambdas) k^2 +(2-\KB) (1+\lambdas) \mu^2}
\label{eq_chi_X_P_Y}
\\
P_\chi  =& \sqrt{\frac{\KB(2-\KB)}{\KB k^2  + (2-\KB) \mu^2}} \frac{\mu k}{\Qcal_0}\left[  \frac{2(2-\KB) \Qcal_0 }{\KB} X  + P_X  \right] 
- \frac{1}{\Qcal_0} \frac{ (2-\KB) (1+\lambdas) \mu^2  }{ (2+\KB \lambdas)k^2 + (2-\KB) (1+\lambdas) \mu^2 }  P_Y  
\\
\alpha =& Y +  \sqrt{\frac{\KB(2-\KB)}{\KB k^2  + (2-\KB) \mu^2}} \frac{\mu k}{\Qcal_0} 
 \left[ \frac{\Qcal_0}{\KB k^2} X + \frac{1}{2} \frac{P_X}{ (2+\KB \lambdas)k^2 + (2-\KB) (1+\lambdas) \mu^2 } \right]
\\
P_{\alpha}  &=  P_Y 
\end{align}
\end{widetext}
where we have also  defined
\begin{align}
\mu^2 \equiv \frac{2 \Kcal_2 \Qcal_0^2 }{2 - \KB}
\end{align}
This gives a Hamiltonian  density
\begin{align}
\tilde{\Hcal} &= \frac{1}{4} |P_X|^2 + \left(c_s^2 k^2 + \Mcal^2\right) |X|^2 
\nonumber 
\\
& + \frac{(2-\KB)^2 \lambdas}{16 \KB \Kcal_2 } \frac{1 - \frac{k_*^2}{k^2}}{  c_s^2 k^2 + \Mcal^2 } |P_Y|^2
\label{Ham_tilde}
\end{align}
where
\begin{align}
k_*^2 &=  \frac{1+\lambdas}{\lambdas} \mu^2
\label{eq_k_star}
\end{align}
We see then that the system can be cast in terms of two decoupled fields, $X$ and $Y$, with canonical momenta $P_X$ and $P_Y$ respectively,
and each field corresponds to one of the normal modes in \eqref{eq_dispersion}. Specifically,
the field $X$ propagates the massive modes in \eqref{eq_dispersion} while the field $Y$ corresponds to the nonpropagating $\omega=0$ modes.

\section{Discussion}
\label{Sec_discussion}
One notices that the sign of the $|P_Y|^2$ term in \eqref{Ham_tilde}  is not positive definite but rather depends on the relevant wave number $k$ and parameters $\lambdas$ and $k_*$.
Clearly as $k\rightarrow \infty$, $|P_Y|^2$ comes with a positive sign provided $\lambdas>0$, and negative otherwise, which provides an additional condition to the one found for
vector modes in \eqref{vector_constraints}.
Taking both scalar and vector mode conditions on the \AeST parameters we require that 
\begin{align}
& 0 < \KB < 2
\\
& 
 \Kcal_2 > 0
\\
& \lambdas >0
\end{align}
These conditions also imply that $\GN > \Gt$ always.

More generally, when $k> k_*$ defined 
by \eqref{eq_k_star}, the Hamiltonian density is positive while when $k< k_*$, negative Hamiltonian density can occur if the $|P_Y|^2$ term in \eqref{Ham_tilde} becomes
significant. The solutions for $\omega=0$ correspond to $Y = A_0(\vec{k}) t + B_0(\vec{k})$ while $P_Y = A_0(\vec{k})$. Thus the mode which could cause negative 
Hamiltonian densities is the one evolving linearly with $t$.
Such instabilities are likely akin to Jeans-type instabilities and do not cause quantum vacuum instability at low momenta~\cite{GumrukcuogluMukohyamaSotiriou2016}.

As discussed in \cite{Skordis:2020eui}, for a spherically symmetric static source of mass $M$, the transition between the MOND and an oscillatory $\mu$-dominated regime 
occurs at $r_C \sim \left(r_M \mu^{-2}\right)^{1/3}$ where $r_M \sim \sqrt{\frac{\GN M}{a_0}}$ is the MOND scale which signifies the transition between the Newtonian 
and MOND regimes on even smaller distances.  Thus, on observational grounds $\mu^{-1}$ must be larger than $\sim \Mpc$, otherwise, the MOND regime would not occur at the scales of galaxies 
at distances $\sim \kpc$ (for the Milky Way $r_M \sim 8 \kpc$).  A system with a MOND scale of $\sim \Mpc$ occurs if its mass is $\sim 10^{15} M_{\odot}$  which is much larger than
typical masses of bound structures.
Thus, for $\lambdas \ge 1$, the scale $k_*$ is always hidden inside the MOND regime (i.e. $k_* < r_M^{-1}$) so that the 
 negative Hamiltonian does not occur in the GR limit for all systems of interest.

At smaller wave numbers $<r_M^{-1}$, \AeST enters the MOND regime (in which case $\lambdas=0$) which would signify that the $Y$-mode always has a negative Hamiltonian. 
However,  then there exists a higher order term $\sim |\Ycal|^{3/2}/a_0 = |\grad \chi|^3/a_0$ that is not part of the analysis above, and  which may stabilize the system.

To investigate this, we set $\lambdas=0$ in the expansion  \eqref{Fcal_exp} and add the MOND term
\begin{align}
\Jcal_{NL} =  \frac{2\lambdas}{3 (1+\lambdas) a_0} |\Ycal|^{3/2}
\end{align}
 where the presence of $\lambdas$ above is a relic of its influence on the observed value of Newton's constant in the strong gravity regime
so that
\begin{equation}
\Fcal =  (2-\KB)  \Jcal_{NL}(\Ycal)  - 2 \Kcal_2  \left(\Qcal - \Qcal_0\right)^2 + \ldots
\label{Fcal_exp}                                                                                                                                                                                  
\end{equation}

 With this, the scalar mode action turns into
\begin{align}
S^{(\scal,new)} = S^{(\scal)} - (2-\KB) \int d^4x  \Jcal_{NL}
\end{align}
where $S^{(\scal)}$ is given by \eqref{realspacescalars}. Since $\Jcal_{NL}$ does not contain any time derivatives, the canonical momenta from section \ref{Sec_Ham} remain the same.
Thus the Hamiltonian analysis of the previous sections follows through so that the deconstrained Hamiltonian is
\begin{align}
\tilde{\Hcal}^{(new)} &=  \tilde{\Hcal} +  \frac{2(2-\KB)\lambdas}{3 (1+\lambdas) a_0} |\grad\chi|^3
\label{Ham_tilde_new}
\end{align}
where $\tilde{\Hcal}$ is given by \eqref{Ham_tilde} (with $\lambdas=0$) and $\chi$ is given by \eqref{eq_chi_X_P_Y}. 

Observe that the nonlinear MOND term above comes with a positive sign and also, it will dominate $\tilde{\Hcal}^{(new)} $ for large $\chi$. Thus,
it is suggestive that the MOND term may make $\tilde{\Hcal}^{(new)} $  to be bounded from below. Indeed, that turns out to be the case for wave numbers $ k > \mu$, as we show in
detail in Appendix \ref{Append}. At smaller wave numbers $k< \mu$ the MOND term is not sufficient to make the Hamiltonian bounded from below, however,
that is the regime where the Minkowski approximation is expected to break down and expanding on FLRW (or even more specifically de Sitter) is more appropriate.

\section{Conclusion}
\label{Sec_conclusion}
We have expanded the action of a newly proposed \AeST theory~\cite{Skordis:2020eui} which has a MOND limit relevant for galactic systems and $\Lambda$CDM limit relevant for cosmology, to second order on Minkowski spacetime. We have identified the normal modes of the fluctuations and shown that
 the propagating vector modes are massive with mass given by \eqref{Mass_of_modes} and speed of sound
~\footnote{This is not to say that they propagate with the speed of light.
We have defined $c_s^2$ as the coefficient of $k^2$ in the dispersion relation.
Only in the limit $k\rightarrow \infty$ does the speed of sound equal to the propagation speed.} equal to the speed of light 
while the propagating scalar modes  also have the same mass, \eqref{Mass_of_modes}, and speed of sound given by \eqref{speed_of_sound}. We identified in addition, 
nonpropagating scalar modes with
dispersion relation $\omega=0$. We computed the deconstrained Hamiltonian of the scalar modes of \AeST on this spacetime and via a canonical transformation have shown that it 
corresponds to a massive particle corresponding to the massive normal mode, and a massless particle corresponding to the mode $\omega=0$. The latter may lead to
negative Hamiltonian densities for wave numbers $k<k_*$ given by \eqref{eq_k_star}. However, as was discussed above, $k_*\lesssim \Mpc^{-1}$ so that such instabilities
do  not occur in the GR limit of the \AeST for all systems of interest. Furthermore, the nonlinear MOND term creates a nontrivial minimum in the Hamiltonian density, so that
it remains bounded from below for all wave numbers $k>\mu$. Performing the same analysis on de Sitter space is left for future investigation.

\section*{Acknowledgments}

C. S. acknowledges support from the European Research Council under the European Union's
Seventh Framework Programme (FP7/2007-2013) / ERC Grant Agreement No. 617656 ``Theories
 and Models of the Dark Sector: Dark Matter, Dark Energy and Gravity''  and from the European Structural and Investment Funds
and the Czech Ministry of Education, Youth and Sports (MSMT) (Project CoGraDS--CZ.02.1.01/0.0/0.0/15003/0000437). T. Z. 
is supported by the Grant Agency of the Czech Republic GACR Grant No. 20-28525S.

\appendix

\section{Canonical momenta and Hamiltonian in Fourier space}
\label{Def_Fourier_Ham}
Consider the  position variable $X(t,\vec{x})$ with Fourier transform
\begin{align}
X(t,\vec{x}) = \int \frac{d^3k}{(2\pi)^3} e^{i\vec{k}\cdot\vec{x}} X_{\vec{k}}(t)
\end{align}
We assume that $X(t,\vec{x})$ is real which imposes $ X^*_{\vec{k}}(t) =  X_{-\vec{k}}(t)$.
The canonical momentum conjugate to  $X(t,\vec{x})$ is defined by
\begin{align}
P(t,\vec{x}) \equiv \frac{\delta S}{\delta \dot{X}(t,\vec{x})}
\end{align}
where $S = \int dt d^3x \Lcal(t,\vec{x})$ is the action expressed in terms of real space fields $X(t,\vec{x})$ and their time derivatives. We may also express $S$ in Fourier space
as $S = \int dt \frac{d^3k}{(2\pi)^3}  \tilde{\Lcal}(t,\vec{k})$ and define the  canonical momentum conjugate to $X_{\vec{k}}(t)$ as
\begin{align}
P_{\vec{k}}(t) \equiv (2\pi)^3 \frac{\delta S}{\delta \dot{X}^*_{\vec{k}}(t)} 
\end{align}
The two canonical momenta thus defined, $P(t,\vec{x})$ and $P_{\vec{k}}(t)$, form a Fourier transform pair. The Hamiltonian then in Fourier space follows as
\begin{align}
H = \int  \frac{d^3k}{(2\pi)^3} \left[ \frac{1}{2} \left( \dot{P}_{\vec{k}} X_{\vec{k}}^* + \dot{P}^*_{\vec{k}} X_{\vec{k}} \right) -  \tilde{\Lcal}(t,\vec{k}) \right]
\end{align}
Note that, 
\begin{align} 
\frac{\delta }{\delta \dot{X}^*_{\vec{k}}(t)}\int d^3q \frac{1}{2} |\dot{X}_{\vec{k}}|^2 
 =& \frac{1}{2} \int d^3q  \bigg[ \delta^{(3)}(\vec{k} + \vec{q}) \dot{X}_{\vec{k}} 
\nonumber
\\
\ \ \ \ 
& + \delta^{(3)}(\vec{k} - \vec{q})  \dot{X}_{\vec{k}}
\bigg] 
\nonumber
\\
=& \dot{X}_{\vec{k}} 
\end{align}
which is useful for evaluating expressions.

\section{The nonlinear MOND term can make the Hamiltonian bounded from below}
\label{Append}
Here we assess the potential of the nonlinear MOND term to make the Hamiltonian bounded from below. The dependence of the MOND term on
$|\grad \chi| = \sqrt{|\grad \chi|^2}$ means that we cannot use Fourier methods: the MOND term is inherently non-Fourier expandable. This is an issue which pertains all field-based realizations of the MOND proposal and it is likely that a future model may be able to solve this by providing a different type of MOND term which is Fourier expandable. 
Nevertheless, we proceed using a naive Fourier space calculation followed by a robust real space calculation to show that the MOND term makes the Hamiltonian bounded from below.

Our conventions in this appendix is that fields with a $\vec{k}$ (or similar) are in Fourier space and fields without, or with $(\vec{x})$ argument are in real space, so that for example,
\begin{align}
\chi(\vec{x}) =& \int \frac{d^3k}{(2\pi)^3} e^{i\vec{k}\cdot\vec{x}} \chi_{\vec{k}} 
\\
\chi_{\vec{k}} =& \int d^3x e^{-i\vec{k}\cdot\vec{x}} \chi(\vec{x})
\end{align}

\subsection{The naive Fourier space based calculation}
Although the MOND term is not strictly speaking Fourier expandable, let us make the naive assumption that in Fourier space we can express 
this as $|\grad \chi(\vec{x})| \rightarrow k \chi_{\vec{k}}$. Then
the Hamiltonian density \eqref{Ham_tilde_new} becomes
\begin{align}
\tilde{\Hcal}^{(new)}_{\vec{k}} =& \frac{1}{4} |P_{X,\vec{k}}|^2 +  |W_{\vec{k}}|^2 - |Z_{\vec{k}}|^2
\\
&
 + \frac{2(2-\KB)\lambdas}{3 (1+\lambdas) a_0} k^3 \left( |\chi_{\vec{k}}|^2 \right)^{3/2}
\end{align}
where we have introduced the new variables which are rescaled versions of the old ones:
\begin{align}
W_{\vec{k}} =& \sqrt{ c_s^2 k^2 + \Mcal^2  } X_{\vec{k}}
\\
Z_{\vec{k}} =& \frac{1}{2\sqrt{2}} \frac{\Mcal}{k \sqrt{ c_s^2 k^2 + \Mcal^2  }} P_{Y,\vec{k}}
\end{align}
 This is not a canonical transformation but it is sufficient for our purpose to determine whether $\tilde{\Hcal}^{(new)}_{\vec{k}}$ has a minimum. In these variables we have that
\begin{align}
\chi_{\vec{k}} =& G_{W,\vec{k}} W_{\vec{k}} +  G_{Z,\vec{k}} Z_{\vec{k}} 
\end{align}
where the kernels $ G_{W,\vec{k}}$ and $  G_{Z,\vec{k}}$ are given by
\begin{align}                                                                                                                                                
G_{W,\vec{k}} =& \frac{\sqrt{\frac{\KB}{2}  c_s^2 k^2  +  \Mcal^2 } }{  k  \sqrt{2-\KB} \sqrt{\Mcal^2 + c_s^2 k^2}}
\\
G_{Z,\vec{k}} =& \frac{\sqrt{2} c_s^2}{2\Mcal}  \frac{k}{\sqrt{\Mcal^2 + c_s^2 k^2}}  
\end{align} 
The extrema of the Hamiltonian are at $P_{X,\vec{k}} = 0$ and according to the conditions
\begin{align}
 W_{\vec{k}} + \frac{(2-\KB)\lambdas}{ (1+\lambdas) a_0} k^3  |\chi_{\vec{k}}|   G_{W,\vec{k}} \chi_{\vec{k}}    &= 0
\label{W_k_chi}
\\
-Z_{\vec{k}} + \frac{(2-\KB)\lambdas}{ (1+\lambdas) a_0} k^3  |\chi_{\vec{k}}|   G_{Z,\vec{k}} \chi_{\vec{k}}   &= 0 
\end{align}
The trivial extremum is at $P_{X,\vec{k}}  = W_{\vec{k}}  = Z_{\vec{k}} = 0$ which corresponds to the saddle point found without the MOND term. However, the MOND term can introduce 
a nontrivial extremum which we now find. Combining the two conditions above yields,
\begin{align}
  G_{Z,\vec{k}}  W_{\vec{k}} + G_{W,\vec{k}} Z_{\vec{k}}  = 0
\end{align}
so that
\begin{align}
\chi_{\vec{k}} =&  -\left(G_{Z,\vec{k}}^2  G_{W,\vec{k}}^{-1} - G_{W,\vec{k}} \right) W_{\vec{k}}
\end{align}
We insert the above condition into \eqref{W_k_chi} so that assuming $ W_{\vec{k}}\ne 0$ we find 
\begin{align}
 S_G \frac{(2-\KB)\lambdas}{ (1+\lambdas) a_0} k^3  |W_{\vec{k}}|  
  &= \frac{ G_{W,\vec{k}}^{-1}}{\left(    G_{Z,\vec{k}}^2  G_{W,\vec{k}}^{-1}   -G_{W,\vec{k}} \right)^2 }
\end{align}
where $S_G = sign[ G_{Z,\vec{k}}^2  G_{W,\vec{k}}^{-1}   -G_{W,\vec{k}} ]$. Since $ G_{Z,\vec{k}}^2  G_{W,\vec{k}}^{-1}   -G_{W,\vec{k}} = 
 \frac{k^2-\mu^2}{(2-\KB) \mu^2 k^2 } G_{W,\vec{k}}^{-1} $, we have that 
$S_G = Sign(k^2-\mu^2)$. Hence, a nontrivial  extremum exists if and only if $k>\mu$.

Completing the calculation, we find
\begin{align}
  W_{\vec{k}}  &=  \frac{\sqrt{2-\KB} (1+\lambdas) a_0 \mu^4 }{\lambdas (k^2 - \mu^2)^2} \frac{\sqrt{\frac{\KB}{2}  c_s^2 k^2  +  \Mcal^2 } }{ \sqrt{\Mcal^2 + c_s^2 k^2}} e^{i\theta_{\vec{k}}}
\end{align}
where $\theta_{\vec{k}}$ is an arbitrary phase, and this leads to 
\begin{align}
\chi_{\vec{k}} =&- \frac{ (1+\lambdas) a_0 \mu^2 }{\lambdas k (k^2 - \mu^2)}  e^{i\theta_{\vec{k}}}
\label{chi_naive}
\end{align}
and so we find that the Hamiltonian density at the nontrivial extremum takes the form
\begin{align}
\tilde{\Hcal}^{(new)}_{\vec{k}} &=  - \frac{(2-\KB) (1+\lambdas)^2 a_0^2 \mu^6 }{3\lambdas^2 (k^2 - \mu^2)^3} 
\label{Hcal_nontrivial}
\end{align}
This is of course negative but that is not an issue and in fact expected. The question is whether $\tilde{\Hcal}^{(new)}_{\vec{k}}$ is a minimum. 
To determine this, we evaluate the Hessian (matrix of second derivatives) at the nontrivial extremum. It is sufficient to consider the subspace spanned by $W_{\vec{k}}$ 
and $Z_{\vec{k}}$ only. 
Letting $A_k = \frac{3(2-\KB)\lambdas}{2(1+\lambdas) a_0} k^3 $ we find
\begin{align}
\Ccal  = \begin{pmatrix}
  1 + A_k  G_{W,\vec{k}}^2  |\chi_{\vec{k}}|
  &
A_k G_{W,\vec{k}} G_{Z,\vec{k}}   |\chi_{\vec{k}}|
\\
A_k G_{W,\vec{k}} G_{Z,\vec{k}}   |\chi_{\vec{k}}|
  & - 1 +  
A_k  G_{Z,\vec{k}}^2  |\chi_{\vec{k}}|
\end{pmatrix}
\end{align}
which is symmetric, and thus, has real eigenvalues. We are interested in the sign of the eigenvalues. The two eigenvalues $\sigma_{\pm}$ obey the characteristic polynomial
\begin{align}
\sigma_{\pm}^2 - ({\rm tr} \Ccal) \sigma_{\pm} + \det\Ccal = 0
\end{align}
Now, $\sigma_+ + \sigma_- =  {\rm tr} \Ccal$ and $\sigma_+ \sigma_- = \det\Ccal $ which explicitly gives
\begin{align}
\sigma_+ + \sigma_- =&  A_k  \left(G_{W,\vec{k}}^2  +  G_{Z,\vec{k}}^2 \right) |\chi_{\vec{k}}| 
\\
\sigma_+  \sigma_- =&   \frac{1}{2} 
\end{align}
Thus both eigenvalues are positive and the \eqref{Hcal_nontrivial} is a minimum (provided that $k>\mu$ so that it exists).


\subsection{Real space calculation}
We now address the same issue in real space. There,
 $\chi$ is given by
\begin{align}
\chi(\vec{x}) =& \int d^3y \left[  G_W(\vec{x}-\vec{y})W(\vec{y}) +  G_Z(\vec{x}-\vec{y})Z(\vec{y}) \right]
\end{align}
where $W(\vec{x})$ and $Z(\vec{x})$ are the real space analog of $W_{\vec{k}} $ and $Z_{\vec{k}} $ respectively, and likewise for the kernels $G_W(\vec{x})$ and $G_Z(\vec{x})$.

Letting
\begin{align}
B_0 =& \frac{(2-\KB)\lambdas}{ (1+\lambdas) a_0}
\end{align}
the deconstrained Hamiltonian density in real space is thus
\begin{align}
\tilde{\Hcal}^{(new)}(\vec{x}) =& \frac{1}{4} P_X^2 +  W^2 - Z^2 + \frac{2}{3} B_0 |\grad\chi|^3
\label{Ham_tilde_new}
\end{align}
To find the extrema, we take the functional derivatives of $\tilde{\Hcal}^{(new)}(\vec{x})$ with respect to $\{P_X, W, Z\}$. This leads to the conditions $P_X = 0$ (as before) and
\begin{align}
 \delta(\vec{y}-\vec{x}) W(\vec{x})   - B_0  G_W(\vec{y}-\vec{x}) \grad\cdot \left( |\grad\chi|\grad \chi \right) =&0,
\label{W_G_W}
\\
 \delta(\vec{y}-\vec{x}) Z(\vec{x})  +   B_0  G_Z(\vec{y}-\vec{x}) \grad\cdot \left(  |\grad\chi| \grad \chi \right) =&0,
\end{align}
which are to be taken to hold in the sense of distributions. 
Combining the above two conditions, leads after manipulation to
\begin{align}
 Z(\vec{x})  =& - \int d^3z \int d^3y   G_Z(\vec{x}-\vec{y}) G^{-1}_W(\vec{y}-\vec{z}) W(\vec{z})  
\end{align}
so that
\begin{align}
\chi  =& \int d^3y G_\chi(\vec{x}-\vec{y}) W(\vec{y})  
\end{align}
where
\begin{align}
G_\chi(\vec{x}) =&
-\frac{c_s^2\sqrt{2-\KB} }{2\Mcal^2}
\int \frac{d^3k}{(2\pi)^3}
 e^{i\vec{k}\cdot\vec{x} }   
\frac{ (k^2 - \mu^2) \sqrt{c_s^2 k^2 + \Mcal^2}}{k  \sqrt{\frac{\KB}{2}  c_s^2 k^2  +  \Mcal^2 }  }
\end{align} 
Consider now the nonlinear term $|\grad \chi|^3 = |\grad \chi| (\grad \chi)^2$ in the Hamiltonian and integrate it by parts to get
\begin{align}
\tilde{\Hcal}^{(new)}(\vec{x}) \rightarrow& \frac{1}{4} P_X^2 +  W^2 - Z^2 
\nonumber
\\
&
- \frac{2}{3} B_0 \chi \grad \cdot \left( |\grad\chi| \grad \chi\right)
\end{align}
Thus, we use \eqref{W_G_W} to eliminate the nonlinear term, leading to
\begin{align}
\tilde{\Hcal}^{(extr)} &=   W^2 - Z^2 - \frac{2}{3} \chi \int d^3y G^{-1}_W(\vec{y}-\vec{x}) W(\vec{y})
\end{align}
Using the kernels, we may then go back to Fourier space to find
\begin{align}
\tilde{\Hcal}^{(extr)} &= - \frac{1}{3} (2-\KB) \mu^2 \int \frac{d^3k}{(2\pi)^3} \frac{k^2}{k^2 - \mu^2} |\chi_{\vec{k}}|^2
\end{align}
Interestingly, inserting the expression \eqref{chi_naive} into the above equation yields the naive Fourier result \eqref{Hcal_nontrivial}.

Provided $k>\mu$ then $\tilde{\Hcal}^{(extr)} < 0$. For the  opposite case $k < \mu$  it would seem that $\tilde{\Hcal}^{(extr)} $ is positive, however, given that
the trivial extremum is a saddle point, this is a contradiction. Rather, for $k < \mu$ the second
 extremum does not exist and the Hamiltonian remains unbounded from below. Back to
the $k>\mu$ case, the nontrivial extremum is in fact a minimum, which is seen due to the fact that the nonlinear term $|\grad\chi|^3$ in \eqref{Ham_tilde_new}
is positive, hence, by increasing $W$ or $Z$, the nonlinear term will dominate and always make the Hamiltonian positive. Thus the MOND term makes the Hamiltonian bounded from below for wave numbers $k> \mu$.

\bibliographystyle{APSunsrt}
\bibliography{references}

\end{document}